# IMPROVING ESTIMATES OF $m\sin i$ BY EXPANDING RV DATASETS

Robert A. Brown[†]
rbrown@stsci.edu

ABSTRACT

We develop new techniques for estimating the fractional uncertainty ($\mathcal{F}$) in the projected planetary mass ($m\sin i$) resulting from Keplerian fits to radial-velocity (RV) datasets of known Jupiter-class exoplanets. The techniques include (1) estimating the distribution of $m\sin i$ using projection, (2) detecting and mitigating chimeras, a source of systematic error, and (3) estimating the reduction in the uncertainty in $m\sin i$ if hypothetical observations were made in the future. We demonstrate the techniques on a representative set of RV exoplanets, known as the Gang of 27, which are candidates for detection and characterization by a future astrometric direct imaging (ADI) mission. We estimate the improvements (reductions) in $\mathcal{F}$ due to additional, hypothetical RV measurements (RVMs) obtained in the future. We encounter and address a source of systematic error, "chimeras," which can appear when multiple *types* of Keplerian solutions are compatible with a single dataset. We find that for $n' = 10$ new, hypothetical RVMs obtained in the last planetary year before 2025, with the same accuracy as the current available RVMs, $\mathcal{F}$ is reduced by ~18%. From there, each plus-one increase in $2\log n' - \log q$, where $q$ is the factor by which RVM measurement uncertainty is reduced, further reduces $\mathcal{F}$ by factor 0.18.

## 1. INTRODUCTION

In the next decade, we can anticipate the development of space telescopes capable of obtaining astrometric direct images (ADIs) and low-resolution spectra of dozens of known radial-velocity (RV) exoplanets. A prime objective is learning their true masses, which are interesting both in themselves, as fundamental astrophysical quantities, and as factors in estimating the atmospheric scale height, which is needed to interpret spectroscopy of the exoplanetary atmosphere. Traub et al. (2016) review the science program of *WFIRST*, which may be the first ADI mission.

Brown (2015) shows that a single ADI measurement of the apparent separation (*s*) between planetary companion and host star can yield an estimate of *sini*, the sine of the inclination angle (*i*) of the planetary orbit. Meanwhile, RV measurements (RVMs) yield estimates of the projected mass (*msini*). Therefore, ADIs and RVMs can be combined to estimate *msini* / *sini* = *m*, the true planetary mass.

Our primary goal in this paper is to formulate and address the following question: What are the potential benefits of possible *future* RVMs, obtained between now and the start of the ADI mission? This question calls for (1) a technique to estimate the fractional uncertainty $\left(\mathcal{F}_{RV}^{msini}\right)$ in *msini* from fitting an RVM dataset ($\mathcal{D}$), and (2) an analysis of how $\mathcal{F}_{RV}^{msini}$ improves when hypothetical RVMs—possibly with increased measurement precision—are combined with the real dataset ($\mathcal{D}_0$).

---
[†] 1017 N. Cowboy Canyon Drive, Green Valley, AZ 85614

The RVM datasets currently available to the public are largely limited to those listed in the discovery papers of RV planets, which in many cases were published many years ago. Therefore, the reader should appreciate that while this paper develops and demonstrates new statistical methods, we can currently demonstrate those methods only with the incomplete RV data currently available to the public.

## 2. RV DATASETS AND ORBITAL SOLUTIONS

In our treatment, the elements of an RV orbital-solution vector (***p***) are:

$$\boldsymbol{p} = (a, e, \omega_p, t_0, msini, V, \dot{V}) \equiv \mathcal{A}[\mathcal{D}], \qquad (1)$$

where *a* is the semimajor axis, *e* is the eccentricity, $\omega_p$ is the planetary argument of periapsis, $t_0$ is an epoch of periapsis passage, *V* is the constant RV of the center of mass, $\dot{V}$ is its radial acceleration, and $\mathcal{A}$ is the process of least-squares fitting regarded as a function of $\mathcal{D}$, which is a dataset of RVMs in the form:

$$\mathcal{D} = \left((t_1, u_1, \delta u_1), (t_2, u_2, \delta u_2), \ldots (t_n, u_n, \delta u_n)\right), \qquad (2)$$

where *n* is the number of RVMs in the dataset, *t* is the epoch of an RVM, and *u* and $\delta u$ are the value and uncertainty of an RVM.

When $\mathcal{A}$ is the usual least-squares fit with weights $1/\delta u^2$, the RV orbital fits are often poor according to the $\chi^2$ metric, presumably due to systematic RV "stellar noise," such as caused by star spots on a rotating star. Therefore, we proceed in this paper using an *unweighted* least-squares fitting process, $\mathcal{A}'$ (all weights assumed to be equal).

We note that the projected mass, $msini \equiv m \sin i$, is the only fitted parameter in ***p*** that relates to the planetary mass. Here, *i* is the unknown inclination angle of the planetary orbit.

# 3. NEW TECHNIQES FOR ESTIMATING $\mathcal{F}_{RV}^{msini}$

To estimate the distribution of the statistic $\mathcal{F}_{RV}^{msini}$ from a dataset $\mathcal{D}$, we employ two Monte Carlo (MC) techniques in series: "projection" and "bootstrap."

Brown (2004) describes MC projection for RVM datasets, following the recipe in §15.6 of Press et al. (2007).

In the current application, projection calls for drawing a large random sample from the population of *msini*. This draw is achieved by fitting RV orbits to each of $N = 10,000$ "jiggled" datasets $\mathcal{D}_j$, each of which is statistically equivalent to the real dataset $\mathcal{D}_0$:

$$\mathcal{D}_j = \mathcal{J}[\mathcal{D}_0] = \left((t_1, uu_1, \delta u_1), (t_2, uu_2, \delta u_2), \ldots (t_n, uu_n, \delta u_n)\right), \quad (3)$$

where *uu* is a normal random variate of mean *u* and standard deviation $\delta u$, and $\mathcal{J}$ is the jiggling process regarded as a function. We obtain $\boldsymbol{p}_j$ by an unweighted least-squares fit, $\mathcal{A}'$, to the dataset $\mathcal{D}_j$. We then obtain the sampled values $msini_j$ by extracting the fifth element of each of the $N$ parameter vectors $\boldsymbol{p}_j$, guided by Equation (1). The result is the projection sample $\{msini_j\}$, for $j = 1$ to $N$.

Ephron (1979) first described our second MC technique, the bootstrap, which estimates the distribution of a statistic without requiring the support of either a parameterized model or the assumption of a normal distribution. The bootstrap operates directly on a sample, which in this case is $\{msini_j\}$, with cardinality 10,000.

In the current application, the desired statistic is $\mathcal{F}_{RV}^{msini}$:

$$\mathcal{F}_{RV}^{msini} \equiv \frac{\sigma_{msini}}{\mu_{msini}}, \quad (4)$$

where $\mu_{msini}$ and $\sigma_{msini}$ are the mean and standard deviation of the population *msini*. Thus, we can compute an estimate of $\mathcal{F}_{RV}^{msini}$ from the mean and standard deviation of the sample $\{msini_j\}$. Further, we can create $N$ "resamples" or "bootstrap samples" by randomly selecting $N$ elements of $\{msini_j\}$ *with replacement*. We compute the statistic $\mathcal{F}_{RV}^{msini}$ for each resample, creating a grand sample $\{\mathcal{F}_{RV,k}^{msini}\}$, for $k = 1$ to $N$. The mean of the $N$ values in $\{\mathcal{F}_{RV,k}^{msini}\}$ is our grand estimate of $\mathcal{F}_{RV}^{msini}$ for the population, which we call $\overline{\mathcal{F}}_{RV}^{msini}$.

## 4. PREPARING HYPOTHETICAL DATASETS ($\mathcal{D}_{\text{hypo}}$)

Here, we prepare a hypothetical dataset ($\mathcal{D}_{\text{hypo}}$), which is the union of the real dataset $\mathcal{D}_0$ and a synthetic dataset $\mathcal{D}_{\text{syn}}$:

$$\mathcal{D}_{\text{hypo}} = \mathcal{D}_0 \cup \mathcal{D}_{\text{syn}}. \tag{5}$$

In the current treatment, we assume that $\mathcal{D}_{\text{syn}}$ comprises $n'$ new RVMs spread evenly over the timespan of one exoplanetary year immediately prior to the assumed start of the ADI mission on 1 January 2025 (i.e., $t_{\text{ADI}}$ = Julian day 24660676.5). Given the planetary period $P$, the epochs of the synthetic RVMs are therefore

$$t'_i = t_{\text{ADI}} - P\left(1 - \frac{i-1}{n'}\right), \tag{6}$$

for $i = 1$ to $n'$. The synthetic dataset is

$$\mathcal{D}_{\text{syn}} = \left((t'_1, u'_1, \delta'_1), (t'_2, u'_2, \delta'_2), \ldots (t'_{n'}, u'_{n'}, \delta'_{n'})\right), \tag{7}$$

where $\delta'_i = q\,\delta_{\text{RMS}}$, where $\delta_{\text{RMS}}$ is the root mean square of the uncertainties in the RVMs contained in $\mathcal{D}_0$, and where

$$u'_i = V + \dot{V}(t'_i - t_0) + \sqrt{G}\,\frac{\left(\cos(v(t'_i) + \omega_s) + e\cos\omega_s\right)}{\sqrt{a(1-e^2)(m+m_s \approx m_s)}}\, m\sin i, \tag{8}$$

where $a$ is the semimajor axis, $e$ is the eccentricity, $\omega_s$ is the argument of periapsis of the star, $m_s$ is the stellar mass, $G$ is the gravitational constant, and the true anomaly $v(t)$ is the root of the equation

$$\tan\frac{v}{2} = \sqrt{\frac{1+e}{1-e}}\,\tan\frac{E}{2}, \tag{9}$$

in which the eccentric anomaly $E$ is the root of Kepler's Equation,

$$E - e\sin E = M, \tag{10}$$

and the mean anomaly $M$ is

$$M = 2\pi\left(\frac{t'_i - t_0}{P}\right). \tag{11}$$

In Equations (6–11), we adopt the orbital elements in $p_0$, which is the RV solution determined by fitting the real dataset $\mathcal{D}_0$, using an unweighted least-squares fit, $\mathcal{A}'$. As a result, the addition of hypothetical RVMs does not substantively change the orbital solution, but it does increase in accuracy by adding supportive information.

In this paper, as a demonstration, we treat five particular cases of hypothetical datasets, as listed in Table 1. Note that Case A assumes *no* additional, synthetic RVMs, and uses only the RVMs in the real dataset $\mathcal{D}_0$.

Table 1. Demonstration cases of $n'$ hypothetical new RVMs obtained in the run-up to the ADI mission.

| Case | $n'$ | $q$ |
|---|---|---|
| A | 0 | ... |
| B | 10 | 1.0 |
| C | 10 | 0.1 |
| D | 100 | 1.0 |
| E | 100 | 0.1 |

Notes. For each planet, the standard deviations $\delta$ of the synthetic RVMs are equal to $q$ times the root-mean-square errors in the real $\mathcal{D}_0$.

# 5. RESULTS FOR $\mathcal{F}_{RV}^{msini}$

Here, we present our results for $\mathcal{F}_{RV}^{msini}$ for Cases A–E.

Our input catalog is the "Gang of 27," which includes all known RV exoplanets that are (1) single-planet systems, (2) hosted by single stars, (3) possibly observable by the ADI mission with an assumed inner working angle IWA = 0.1″, and (4) offer a publically available RV dataset. (See Table A.1 and Appendices I and II.)

We prepare columns 6–10 in Table 2 using the recipes for MC projection and bootstrap described in §3. Each entry $\bar{\mathcal{F}}_{RV}^{msini}$ is the mean of 10,000 realizations of $\mathcal{F}_{RV}^{msini}$ computed from jiggled datasets, each one fitted for a value of *msini*, followed by 10,000 bootstrap resamples of *msini*, from each of which we compute one estimate of the mean and one estimate of the standard deviation of *msini*. From each such pair of values we calculate one value of $\mathcal{F}_{RV}^{msini}$ by Equation (4). Those steps produce a sample of 10,000 values of $\mathcal{F}_{RV}^{msini}$. Our grand estimate for Table 2 is $\bar{\bar{\mathcal{F}}}_{RV}^{msini}$, which is the mean of this sample of $\mathcal{F}_{RV}^{msini}$.

As explained in §6, the results in red typeface in Table 2 are affected by chimeras. The 16 affected exoplanets are excluded from the final analysis.

The three columns headed by a reference to "*msini*" are a reality check on our Keplerian fits of jiggled datasets. The third column is the value of *msini* listed in the www.exoplanets.org catalog on 1 March 2014. The fourth column is the mean of *msini* from the 10,000 fits to jiggled datasets. These are purely Case A results, involving no synthetic RVMs. The fifth column is the fractional deviation between the third and fourth columns. In other words, it is a comparison between our Keplerian fits and those published by the RV practitioners.

In the fifth column, consider at first only the 11 values of fractional deviation in black typeface. The average absolute fractional deviation is 1.5%. For these 11 exoplanets, we are therefore confident that we are performing the Keplerian fits correctly as stated. The small differences with the catalog results for *msini* could be explained by any of several factors, including the questions of weighted versus unweighted least-squares fits, the inclusion or not of a customary "stellar jitter" term to lower the chi-square metric (not used here), and the particular suite of Keplerian parameters chosen to fit.

By contrast, most of the 16 exoplanets with red typeface in the fifth column of Table 2 show elevated values of fractional deviation. Those are the exoplanets with chimeras in Case A, as we discuss in §6.

Table 2. Estimates of $\bar{\mathcal{F}}^{msini}_{RV}$, the fractional error in *msini* for the five cases of hypothetical datasets, A–E.

| | Gang of 27 | Catalog *msini* | Case A *msini* | $\frac{\Delta msini}{msini}$ | Case A | B | C | D | E |
|---|---|---|---|---|---|---|---|---|---|
| 1 | GJ 649 b | 0.325 | 0.360 | 0.106 | 0.064 | 0.042 | 0.034 | 0.026 | 0.012 |
| 2 | HD 147513 b | 1.180 | 1.112 | -0.057 | 0.100 | 0.044 | 0.038 | 0.022 | 0.011 |
| 3 | HD 72659 b | 3.174 | 3.684 | 0.161 | 0.543 | 0.960 | 0.671 | 0.019 | 0.010 |
| 4 | HD 70642 b | 1.909 | 1.961 | 0.027 | 0.036 | 0.027 | 0.023 | 0.014 | 0.007 |
| 5 | 7 CMa b | 2.432 | 4.539 | 0.866 | 0.579 | 1.168 | 1.380 | 0.666 | 0.134 |
| 6 | HD 50554 b | 4.399 | 5.078 | 0.154 | 0.091 | 0.034 | 0.018 | 0.014 | 0.004 |
| 7 | GJ 832 b | 0.644 | 4.258 | 5.608 | 1.111 | 0.312 | 0.739 | 1.096 | 1.149 |
| 8 | 14 Her b | 5.215 | 5.151 | -0.012 | 0.006 | 0.005 | 0.005 | 0.003 | 0.002 |
| 9 | GJ 849 b | 0.831 | 0.821 | -0.012 | 0.489 | 2.767 | 2.524 | 1.869 | 1.046 |
| 10 | HD 79498 b | 1.346 | 1.332 | -0.010 | 0.042 | 0.035 | 0.034 | 0.023 | 0.016 |
| 11 | 16 Cyg B b | 1.640 | 1.665 | 0.015 | 0.030 | 0.028 | 0.026 | 0.018 | 0.012 |
| 12 | HD 39091 b | 10.088 | 10.086 | -0.000 | 0.006 | 0.006 | 0.005 | 0.004 | 0.002 |
| 13 | HD 216437 b | 2.168 | 2.173 | 0.002 | 0.022 | 0.020 | 0.018 | 0.013 | 0.007 |
| 14 | HD 220773 b | 1.450 | 2.473 | 0.706 | 1.696 | 1.358 | 0.979 | 0.506 | 0.128 |
| 15 | HD 50499 b | 1.745 | 1.749 | 0.003 | 0.043 | 0.036 | 0.030 | 0.020 | 0.010 |
| 16 | GJ 676 A b | 4.897 | 4.810 | -0.018 | 0.002 | 0.002 | 0.001 | 0.001 | 0.001 |
| 17 | HD 89307 b | 1.791 | 1.770 | -0.012 | 0.066 | 0.046 | 0.042 | 0.028 | 0.018 |
| 18 | epsilon Eri b | 1.054 | 0.953 | -0.096 | 0.170 | 0.058 | 0.062 | 0.036 | 0.027 |
| 19 | HD 154345 b | 0.957 | 0.936 | -0.022 | 0.035 | 0.024 | 0.021 | 0.012 | 0.006 |
| 20 | HD 117207 b | 1.819 | 3.425 | 0.883 | 1.330 | 0.024 | 0.021 | 0.014 | 0.008 |
| 21 | GJ 179 b | 0.824 | 0.815 | -0.012 | 0.059 | 0.051 | 0.046 | 0.030 | 0.014 |
| 22 | HD 10647 b | 0.925 | 0.938 | 0.014 | 0.360 | 0.692 | 0.347 | 0.039 | 0.019 |
| 23 | HD 87883 b | 1.756 | 4.288 | 1.443 | 1.554 | 0.226 | 0.124 | 0.024 | 0.013 |
| 24 | HD 30562 b | 1.333 | 1.400 | 0.051 | 0.113 | 0.624 | 0.190 | 0.493 | 0.020 |
| 25 | GJ 317 b | 1.175 | 1.315 | 0.119 | 0.339 | 0.016 | 0.013 | 0.008 | 0.003 |
| 26 | HD 142022 b | 4.468 | 7.518 | 0.683 | 1.433 | 0.080 | 0.024 | 0.014 | 0.003 |
| 27 | HD 106252 b | 6.959 | 6.641 | -0.046 | 0.017 | 0.013 | 0.012 | 0.009 | 0.006 |

Many of the values of $\bar{\mathcal{F}}^{msini}_{RV}$ in columns 6–10 in Table 2 are anomalous—such as those that are greater than one. Other values seem reasonable, in the range 1–10%. Furthermore, for those planets showing a sequence of reasonable values, the sequence shows monotonic improvement (decline) with the progression of Cases A–E. This is as expected; from the discussion of Table 1 in §7, we expect that the information content of the synthetic datasets increases monotonically in ascending alphabetic order, Case A to Case E.

Our main goal is to estimate the *improvements* in $\mathcal{F}_{RV}^{msini}$ when hypothetical RVMs are added to the real dataset. We measure this improvement by the metric $f$, which is the fractional change in $\mathcal{F}_{RV}^{msini}$ for Cases B–E as compared with Case A:

$$f(\text{B–E}) \equiv \frac{\mathcal{F}_{RV}^{msini}(\text{B–E})}{\mathcal{F}_{RV}^{msini}(\text{A})} \quad . \tag{12}$$

Equation (12) applies separately to any exoplanet in the Gang of 27. As an example, for Case D and HD 89307 b, $f(\text{D}, \text{HD 89307 b}) = 0.028/0.066 = 0.42$. For that particular exoplanet, adding 100 synthetic RVM of the same root-mean-square accuracy as in the real dataset reduces the uncertainty in *msini* by 42%.

# 6. CHIMERAS

RV chimeras are a phenomenon previously described—but not given a name—by Brown (2004), who studied the chimera in HD 72659 b.

Chimeras arise in the projection sample $\{msini_j\}$ when the dataset supports two or more *types* of Keplerian solution. (We borrow the term "chimera" from genetics, where it refers to an organism composed of two or more genetically distinct tissues, such as an organism that is partly male and partly female.) We find that usually one solution type is dominant and centered near the primary solution. The other solution type—the chimera—may exploit a lack of constraint in the dataset, sometimes producing a multi-modal distribution of RV parameters.

As an example, Figures 1–2 show the chimera of epsilon Eri , discovered in the current study.

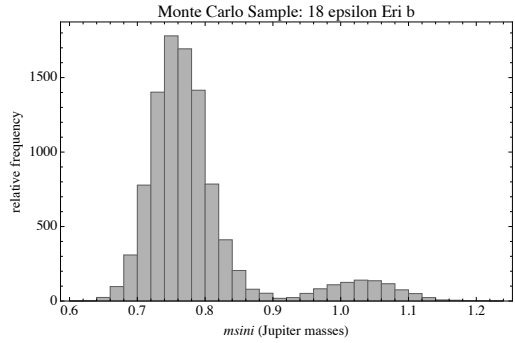

Figure 1. A chimera in the Case-A projection sample of epsilon Eri b. The histogram is based on $N = 10{,}000$ values of *msini* computed by fitting jiggled datasets, as in Equation (3).

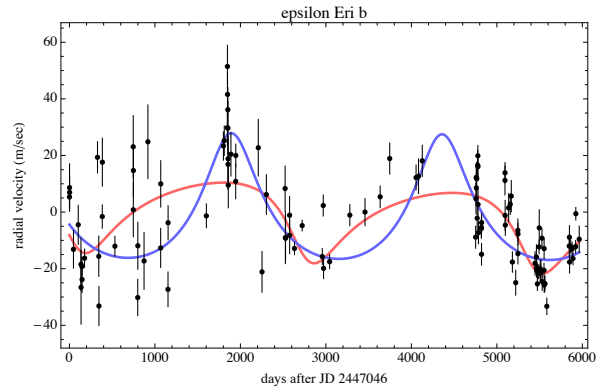

Figure 2. Typical primary (red) and chimeric (blue) solutions to the Case A projection sample for epsilon Eri b. The eccentricity $e$ and *msini* are $e = 0.36$ and $msini = 0.75\ m_{\mathrm{Jupiter}}$ for the primary solution, and $e = 0.32$ and $msini = 1.22\ m_{\mathrm{Jupiter}}$ for the chimeric solution with the greatest deviation from the primary solution.

The skewness metric is useful for detecting outliers and asymmetric distributions, such as those caused by chimeras. Our application of the metric uses the absolute value of skewness, defined as

$$skewness \equiv \left| \frac{\mu_3}{\mu_2^{3/2}} \right|, \qquad (13)$$

where $\mu_i$ is the $i^{th}$ central moment of the distribution $\{msini_j\}$:

$$\mu_i \equiv \left\langle \left( msini - \langle msini \rangle \right)^i \right\rangle. \qquad (14)$$

Table 3. The *skewness* metric computed from the projection samples $\{msini_j\}$.

| | Gang of 27 | Case | | | | |
|---|---|---|---|---|---|---|
| | | A | B | C | D | E |
| 1 | GJ 649 b | 0.441 | 0.053 | 0.015 | 0.019 | 0.026 |
| 2 | HD 147513 b | 0.596 | 0.006 | 0.014 | 0.018 | 0.003 |
| 3 | HD 72659 b | 16.738 | 35.714 | 43.565 | 0.007 | 0.051 |
| 4 | HD 70642 b | 0.013 | 0.011 | 0.015 | 0.032 | 0.008 |
| 5 | 7 CMa b | 2.970 | 2.113 | 2.164 | 17.603 | 16.353 |
| 6 | HD 50554 b | 1.209 | 0.025 | 0.086 | 0.001 | 0.002 |
| 7 | GJ 832 b | 0.947 | 85.780 | 36.786 | 1.140 | 1.407 |
| 8 | 14 Her b | 0.020 | 0.037 | 0.011 | 0.018 | 0.005 |
| 9 | GJ 849 b | 58.610 | 5.998 | 5.559 | 3.596 | 11.040 |
| 10 | HD 79498 b | 0.510 | 0.199 | 0.211 | 0.026 | 0.015 |
| 11 | 16 Cyg B b | 0.065 | 0.003 | 0.030 | 0.017 | 0.010 |
| 12 | HD 39091 b | 0.132 | 0.003 | 0.058 | 0.039 | 0.007 |
| 13 | HD 216437 b | 0.017 | 0.003 | 0.031 | 0.033 | 0.045 |
| 14 | HD 220773 b | 5.692 | 7.658 | 12.598 | 9.567 | 25.674 |
| 15 | HD 50499 b | 0.005 | 0.027 | 0.019 | 0.021 | 0.033 |
| 16 | GJ 676 A b | 0.026 | 0.011 | 0.014 | 0.010 | 0.019 |
| 17 | HD 89307 b | 0.112 | 0.051 | 0.002 | 0.040 | 0.014 |
| 18 | epsilon Eri b | 66.296 | 0.492 | 0.723 | 0.004 | 0.019 |
| 19 | HD 154345 b | 0.111 | 0.038 | 0.023 | 0.049 | 0.012 |
| 20 | HD 117207 b | 2.970 | 0.025 | 0.004 | 0.014 | 0.003 |
| 21 | GJ 179 b | 0.139 | 0.084 | 0.025 | 0.009 | 0.036 |
| 22 | HD 10647 b | 41.009 | 32.221 | 70.945 | 0.036 | 0.026 |
| 23 | HD 87883 b | 2.531 | 22.782 | 22.936 | 0.306 | 0.347 |
| 24 | HD 30562 b | 51.674 | 4.285 | 20.115 | 28.715 | 4.201 |
| 25 | GJ 317 b | 50.547 | 0.005 | 0.019 | 0.020 | 0.013 |
| 26 | HD 142022 b | 5.004 | 1.312 | 0.642 | 0.001 | 0.003 |
| 27 | HD 106252 b | 0.030 | 0.005 | 0.000 | 0.002 | 0.029 |

Note: As in Table 2, red typeface in Case A indicates corruption by chimeras, and the exoplanet is excluded from the final analysis.

We find strong correlations between (1) evidence of chimeras in histograms of $\{msini_j\}$, such as Figure 1, (2) elevated values of the *skewness* of $\{msini_j\}$, as shown by the results in red typeface in Table 3, and (3) bias in the calculation of $\bar{\mathcal{F}}_{RV}^{msini}$, as shown by the results in red typeface in Table 2. On the basis of these correlations—particularly that between items (2) and (3)—we include in the final analysis only the subset of results $\bar{\mathcal{F}}_{RV}^{msini}$ in Table 2 for which the Case-A *skewness* in Table 3 is less than a chosen threshold value, *cutoff* = 0.3).

Our estimator of $f$ in Equation (12) is its mean value, $\langle f \rangle$, computed on the valid subset of 11 exoplanets. Because the value of *cutoff* defines the valid subset, the estimate $\langle f \rangle$ is a function of *cutoff*, too. If *cutoff* is lowered, the cardinality of the valid subset is reduced and $f$ gets noisy. If *cutoff* is sufficiently increased, then chimera-corrupted values of $\bar{\mathcal{F}}_{RV}^{msini}$ are included, and again $f$ gets noisy. The choice for *cutoff* involves a tradeoff, as shown in Figure 3. We choose *cutoff* = 0.3, which minimizes noise towards both limits.

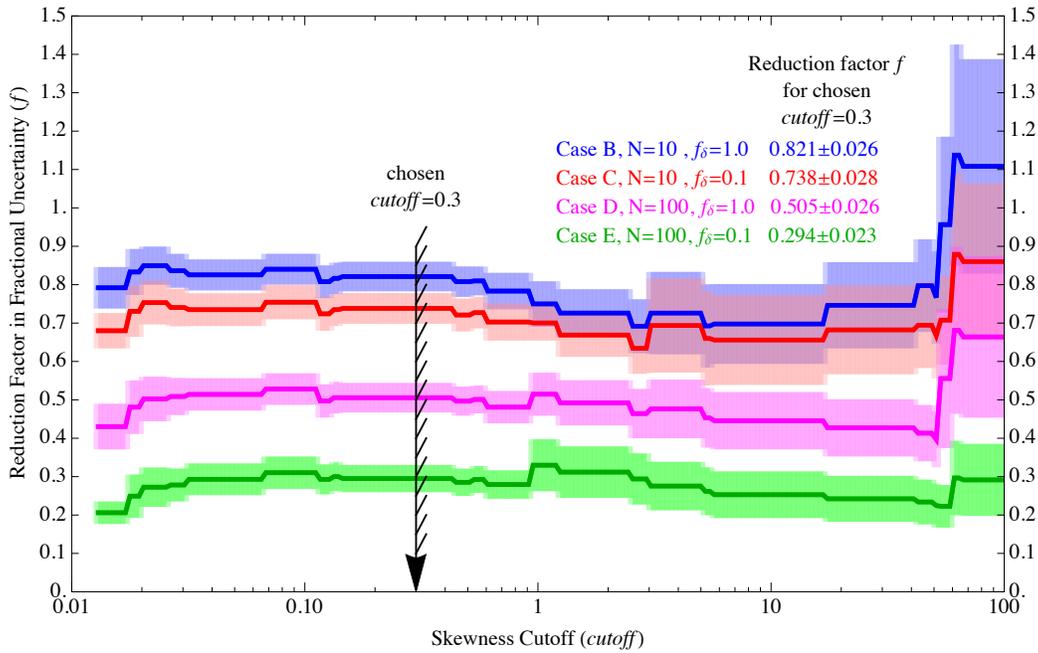

Figure 3. The ordinate, $f$, is the factor by which the fractional uncertainty in *msini*, $\mathcal{F}_{RV}^{msini}$, is changed (decreased) by the new RVMs in hypothetical datasets defined by Cases B–E. The abscissa is *cutoff*, which is the threshold value of the *skewness* metric. The color bands show plus and minus the standard deviation of the mean, $\langle f \rangle$. The combinations of exoplanet and case excluded by the criterion *skewness* < *cutoff* = 0.3 in Case A are indicated by red typeface in Tables 2 and 3. The value *cutoff* = 0.3 was selected to minimize the uncertainty (width of colored bands) and to avoid cimeras to the right and small-sample noise to the left.

# 7. RESULTS FOR $f$

Table 4 and Figure 4 show our final results for $f$, the factor by which $\mathcal{F}$ is improved (reduced) by hypothetical new RVMs that might be obtained in the future. We compute $f$ from Equation (14) using values of $\bar{\mathcal{F}}_{RV}^{msini}$ in Table 2. Only chimera-free RV exoplanets contributed to these results.

The abscissa in Figure 4 is

$$x = 2\log n' - \log q . \tag{15}$$

The blue line in Figure 4 is a least-squares to the data in Table 4, which is the function

$$f(x) = 1.224 - 0.1811x \tag{16}$$

For Case B, involving $n' = 10$ hypothetical RVMs with the same accuracy as the real RVMs, $\mathcal{F}$ is reduced by the factor $f \simeq 0.18$ or ~18% with respect to Case A, which has only real data points. From there, each plus-one increase in $x$—which may be achieved by various combinations of changes in $n'$ and $q$— further reduces $f$ by factor 0.18.

For example, according to Equation (15), $x$ changes by plus one if $n'$ increases by factor $\sqrt{10}$ or if $q$ decreases by factor 10. This confirms the natural expectation that the accuracy of *msini* varies directly as $n'$, the square root of the number of data points, and inversely as the reduction in the measurement errors, $q$.

Table 4. Results for $f$, the reduction factor in fractional uncertainty $\mathcal{F}$, for Cases B–E of hypothetical datasets.

| Case | B | C | D | E |
|---|---|---|---|---|
| $\langle f \rangle$ | $0.821 \pm 0.026$ | $0.738 \pm 0.028$ | $0.505 \pm 0.026$ | $0.294 \pm 0.023$ |
| $x$ | 2. | 3. | 4. | 5. |

Note: Only the 11 chimera-free exoplanets in the Gang of 27 contributed to these results. The quoted uncertainty is the standard deviation of the mean of $f$, which is $\langle f \rangle$.

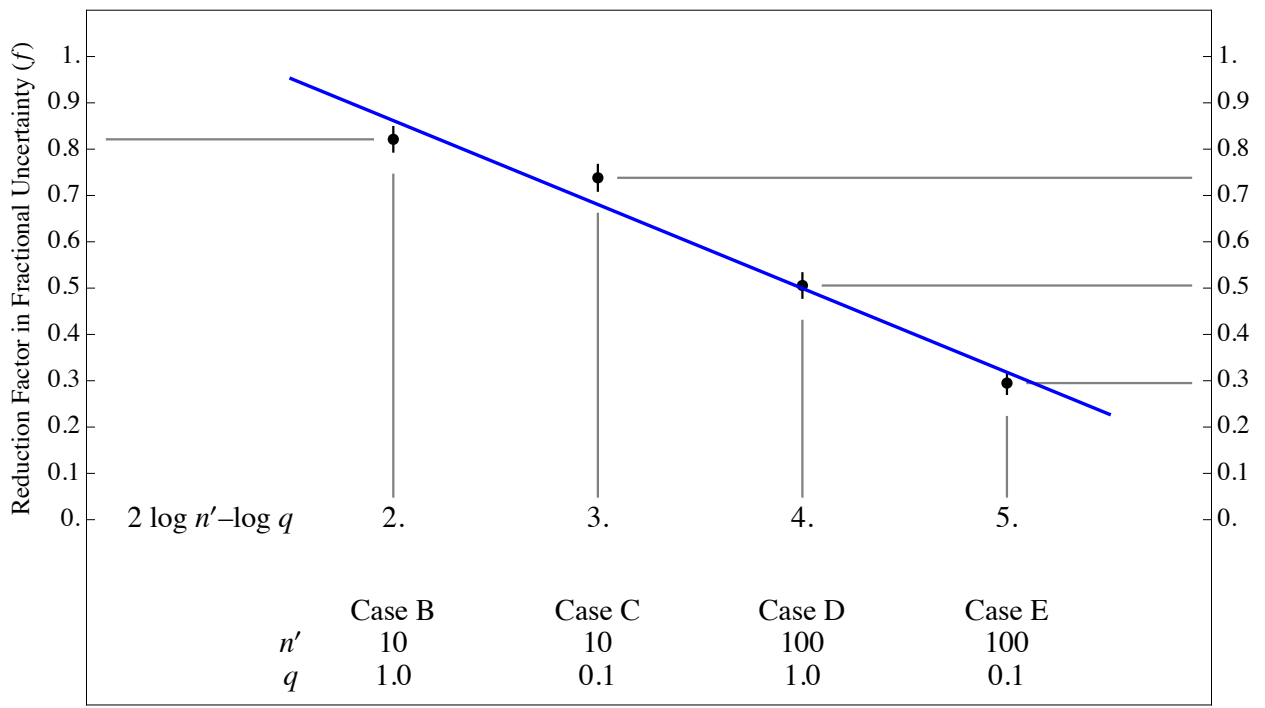

Figure 4. Final results for *f*, from Table 4. Our data confirm that the accuracy of *msini* varies directly as $n'$, the square root of the number of data points, and inversely as the reduction in the measurement errors, $q$.

# 8. CONCLUDING REMARKS

We develop three new techniques in this paper: (1) estimating the distribution of *msini* using Monte Carlo projection, (2) detecting and mitigating chimeras using the metric *skewness*, and (3) estimating the reduction in the uncertainty in *msini* if hypothetical observations are made in the future. We have demonstrated the effectiveness of these techniques on a uniform subset of 11 exoplanets in the Gang of 27.

For the other 16 exoplanets in the Gang of 27, corruption by chimeras introduces systematic error into estimates of *msini*—and no doubt other Keplerian parameters. These errors can be mitigated by more RVMs, as is illustrated by comparing Case A and Case E in Table 3. Only five exoplanets in Case E—the case involving the most new information—show evidence of chimeras. It is possible that RVMs already obtained have eliminated chimeras in many RV exoplanets. Many more datasets and RVMs exist but are off limits, in private hands, and unavailable for research by the scientific public.

By applying these new techniques to new cases of hypothetical datasets, it is now possible to optimize future RV programs in support of future ADI missions, such as *WFIRST*. In fact, an RV practitioner can now estimate the improvement in the accuracy of *msini* due to a newly obtained RVM, in real time, at the telescope.

ACKNOWLEDGEMENTS

We thank Stuart Shaklan, Marc Postman, Wes Traub, and Dave Latham for their support and encouragement of this work over many years. We are grateful to Bob Vanderbei for his advice on statistical issues from time to time. Jason Wright has been helpful with the acquisition of publically available RV datasets for the Gang of 27.

APPENDIX A: THE GANG OF 27

Our input catalog, which we call the "Gang of 27," includes all 27 known RV exoplanets that are (1) single-planet systems, (2) hosted by single stars, (3) possibly observable by the ADI mission with an assumed inner working angle, IWA = 0.1″, and (4) provided with a publically available RV dataset.

To develop this catalog, we start with all 436 RV planets in the catalog at the website www.exoplanets.org on 1 March 2014. In the first cut, we eliminate all but 55 exoplanets on the basis of the criterion $s_{raw} \equiv (1 + e) \, a / d > 0.1″$. The term on the left side of this inequality, which we call the "raw separation," is the maximum possible value of $s$ for an exoplanet at stellar distance = $d$, with semimajor axis = $a$, eccentricity = $e$, but unknown inclination $i$. The Gang of 55 is listed in Table 1.

In the second cut, we eliminate all multiple-planet systems, for simplicity in the orbital solutions. This cut creates the Gang of 28.

In the third cut, we eliminate the one planet that never achieves $s > 0.1″$ despite the fact $s_{raw} > 0.1″$. As explained in the Appendix, we use a new completeness metric—global observational completeness (GOC)—to test the entire Gang of 28 for GOC > 0, and we find that only one of them—kappa CrB b—does not satisfy the GOC criterion.

It is important to recognize that the Gang of 27 is *not* a sample developed for statistical purposes. It is simply a list of RV exoplanets useful for demonstrating our new techniques for estimating $\mathcal{F}_{RV}^{msini}$.

Table A.1. Input catalog: the Gang of 27.

| 1 | 2 | 3 | 4 | 5 | 6 | 12 | 13 |
|---|---|---|---|---|---|---|---|
|   | Gang of 55 | single planet | single star | RV data available | GOC | Gang of 27 |   |
| 1 | HD 24040 b | no | yes | no |   |   |   |
| 2 | GJ 649 b | yes | yes | yes | 0.415 | GJ 649 b | 1 |
| 3 | HD 38529 c | no | yes | no |   |   |   |
| 4 | HD 147513 b | yes | yes | yes | 0.288 | HD 147513 b | 2 |
| 5 | HD 72659 b | yes | yes | yes | 0.0346 | HD 72659 b | 3 |
| 6 | HD 70642 b | yes | yes | yes | 0.465 | HD 70642 b | 4 |
| 7 | GJ 328 b | no | yes | no |   |   |   |
| 8 | 7 CMa b | yes | yes | yes | 0.436 | 7 CMa b | 5 |
| 9 | HD 192310 c | no | yes | yes | | | |
| 10 | HD 204313 d | no | yes | no | | | |
| 11 | HD 50554 b | yes | yes | yes | 0.308 | HD 50554 b | 6 |
| 12 | GJ 832 b | yes | yes | yes | 0.613 | GJ 832 b | 7 |
| 13 | mu Ara b | no | yes | yes | | | |
| 14 | 14 Her b | yes | yes | yes | 0.710 | 14 Her b | 8 |
| 15 | GJ 849 b | yes | yes | yes | 0.794 | GJ 849 b | 9 |
| 16 | HD 79498 b | yes | yes | yes | 0.016 | HD 79498 b | 10 |
| 17 | HD 33636 b | no | yes | no | | | |
| 18 | HD 217107 c | no | yes | no | | | |
| 19 | HD 13931 b | no | yes | no | | | |
| 20 | 16 Cyg B b | yes | yes | yes | 0.103 | 16 Cyg B b | 11 |
| 21 | HD 39091 b | yes | yes | yes | 0.324 | HD 39091 b | 12 |
| 22 | HD 7449 b | no | yes | no | | | |
| 23 | HD 216437 b | yes | yes | yes | 0.160 | HD 216437 b | 13 |
| 24 | HD 220773 b | yes | yes | yes | 0.008 | HD 220773 b | 14 |
| 25 | HD 181433 d | no | yes | yes | | | |
| 26 | kappa CrB b | yes | yes | yes | 0. | | |
| 27 | HD 204941 b | no | yes | no | | | |
| 28 | HD 50499 b | yes | yes | yes | 0.007 | HD 50499 b | 15 |
| 29 | GJ 676 A b | yes | yes | yes | 0.400 | GJ 676 A b | 16 |
| 30 | HD 89307 b | yes | yes | yes | 0.322 | HD 89307 b | 17 |
| 31 | epsilon Eri b | yes | yes | yes | 0.800 | epsilon Eri b | 18 |
| 32 | HD 154345 b | yes | yes | yes | 0.644 | HD 154345 b | 19 |
| 33 | mu Ara c | no | yes | no | | | |
| 34 | HD 169830 c | no | yes | no | | | |
| 35 | 47 UMa c | no | yes | no | | | |
| 36 | HD 134987 c | no | yes | no | | | |
| 37 | 55 Cnc d | no | yes | yes | | | |
| 38 | HD 222155 b | no | yes | no | | | |
| 39 | HD 117207 b | yes | yes | yes | 0.519 | HD 117207 b | 20 |
| 40 | GJ 179 b | yes | yes | yes | 0.829 | GJ 179 b | 21 |
| 41 | HD 128311 c | no | yes | no | | | |
| 42 | HD 10180 h | no | yes | no | | | |
| 43 | HD 10647 b | yes | yes | yes | 0.412 | HD 10647 b | 22 |
| 44 | HD 196885 b | no | no | yes | | | |
| 45 | HD 87883 b | yes | yes | yes | 0.260 | HD 87883 b | 23 |
| 46 | gamma Cep b | no | no | yes | | | |
| 47 | beta Gem b | no | yes | no | | | |
| 48 | 47 UMa b | no | yes | yes | | | |
| 49 | HD 30562 b | yes | yes | yes | 0.190 | HD 30562 b | 24 |
| 50 | HD 190360 b | no | yes | yes | | | |
| 51 | GJ 317 b | yes | yes | yes | 0.334 | GJ 317 b | 25 |
| 52 | HD 142022 b | yes | yes | yes | 0.473 | HD 142022 b | 26 |

| 53 | HD 187123 c | no | yes | no | | | |
| 54 | HD 106252 b | yes | yes | yes | 0.014 | HD 106252 b | 27 |
| 55 | upsilon And d | no | yes | yes | | | |

Notes: GOC > 0 means that the planet, at some time in its orbit, achieves $s >$ IWA $= 0.1''$. In this table, GOC is shown only for the members of the Gang of 28 ("yes" in all columns 3–5). Kappa CrB b is the only planet in the Gang of 28 to be eliminated because GOC $\not> 0$. Columns 4–5 are criteria that, by themselves, eliminate no exoplanets (no single-planet systems with multiple host stars or for which we don't have an RV dataset). In the "Gang of 27" column, blue typeface typeface means the exoplanet is chimera-free and included in the final results.

## APPENDIX B: GLOBAL OBSERVATIONAL COMPLETENESS

Global observational completeness (GOC) is a new observability metric for RV orbits. It serves a real purpose only for the narrow case when $e > 0$, $s_{raw} >$ IWA, and GOC $= 0$. Any planet satisfying those three criteria can be dropped from the input catalog for an ADI mission because the orbit is oriented in such a way that the planet is never observable. In the current study, we eliminate one planet from the Gang of 28: kappa CrB b, making the Gang of 27.

We compute GOC from the orbital solution $p_0$ that results from fitting the "catalog" RV dataset ($\mathcal{D}_0$). If the inclination $i$ were known and combined with $p_0$, we could compute the apparent separation $s$ at any time. We do not know $i$, of course—in fact, we are trying to estimate it. Nevertheless, we know how $i$ is *distributed*, namely uniformly in cos $i$ and with the random variate arccos$(1 - \mathcal{R})$, where $\mathcal{R}$ is a uniform random variate on the range 0–1. That information is enough to compute GOC by Monte Carlo experiment, as follows. First, we draw a random value of time $t$ from a uniform random variate on the range 0–$10^9$ days—in other words, anytime. Second, we draw a value of $i$ from its random variate. Third, we combine each value of $i$ with $p_0$ to compute $s$ and $\Delta mag$ from Equations (8, 20, 21) in Brown (2015). Fourth, we apply the detectability criteria $s >$ IWA and $\Delta mag < \Delta mag_0$, and get a 1 if both criteria are satisfied and get a 0 otherwise. Fifth, we sum the values for the one million trials, and divide by one million. Voila: GOC.

Table 1 shows the values of GOC for all the planets in the Gang of 28.